\documentclass[prb,twocolumn,showpacs,floatfix]{revtex4}
\usepackage{graphicx}
\usepackage{times}
\usepackage{bm}

\def\be{\begin{equation}}
\def\ee{\end{equation}}
\def\bea{\begin{eqnarray}}
\def\eea{\end{eqnarray}}
\def\bse{\begin{subequations}}
\def\ese{\end{subequations}}

\def\be{\begin{eqnarray}}
\def\ee{\end{eqnarray}}

\begin{document}

\title{Anomalous Nernst effect from a chiral $d$-density wave state in
underdoped cuprate superconductors}
\author{Chuanwei Zhang$^{1,2}$}
\author{Sumanta Tewari$^{1,3}$}
\author{Victor M. Yakovenko$^{1}$}
\author{S. Das Sarma$^{1}$}
\affiliation{$^{1}$Condensed Matter Theory Center, Department of Physics, University
ofMaryland, College Park, Maryland 20742, USA\\
$^{2}$Department of Physics and Astronomy, Washington State University,
Pullman, WA 99164 USA\\
$^{3}$Department of Physics and Astronomy, Clemson University, Clemson, SC
29634 USA }

\begin{abstract}
We show that the breakdown of time-reversal invariance, confirmed by the
recent polar Kerr effect measurements in the cuprates, implies the existence
of an anomalous Nernst effect in the pseudogap phase of underdoped cuprate
superconductors. Modeling the time-reversal-breaking ordered state by the
chiral $d$-density-wave state, we find that the magnitude of the Nernst
effect can be sizable even at temperatures much higher than the
superconducting transition temperature. These results imply that the
experimentally found Nernst effect at the pseudogap temperatures may be due
to the chiral $d$-density wave ordered state with broken time-reversal
invariance.
\end{abstract}

\pacs{74.72.-h, 72.15.Jf, 72.10.Bg}
\maketitle

\section{Introduction}

Even after two decades of intensive research, the physics of the high
temperature cuprate superconductors is as elusive as ever \cite{Lee}. The
principal mystery surrounds the underdoped regime, which evinces a
well-formed quasiparticle gap even at temperatures well above the
superconducting transition temperature, $T_{c}$.
The recent observation of a non-zero polar Kerr effect (PKE) in the
underdoped YBCO \cite{Xia1}, which demonstrates macroscopic time-reversal
(TR) symmetry breaking in the pseudogap phase, is a step forward in solving
the pseudogap puzzle. The PKE appears roughly at the same temperature, $%
T^{\ast }$, where the pseudogap develops \cite{Xia1}. Near optimum doping,
the PKE appears at a temperature \textit{below} $T_{c}$, consistent with the
existence of a zero temperature quantum phase transition under the
superconducting dome. This observation suggests that the TR symmetry
breaking and the pseudogap in the cuprates may have the same physical
origin, which is also unrelated to the $d$-wave superconductivity itself.
Similar conclusion was also reached eariler by muon spin rotation
experiments \cite{Sonier}. In this work we predict the existence of an
anomalous Nernst effect associated with the TR symmetry breaking which
should be present along with the observed PKE in the underdoped cuprates.
Our results demonstrate that the existence of anomalous Nernst effect at
temperatures as high as the pseudogap temperatures (see below), where the
vortex excitations of the superconductor are unlikely to be present, may
imply an ordered state with broken TR symmetry in the pseudogap regime of
the underdoped cuprates.

It was proposed earlier \cite{Chakravarty01,Chakravarty04} that the $%
id_{x^{2}-y^{2}}$ density-wave (DDW) state may be responsible for the
pseudogap behavior in the underdoped cuprates. In real space, the order
parameter for this state consists of orbital currents along the bonds of the
two dimensional square lattice of Copper atoms. Since the currents circulate
in opposite directions in any two consecutive unit cells of the lattice, the
total orbital current averages to zero, and the macroscopic TR symmetry
remains unbroken. Recently, it was shown that the admixture of a small $%
d_{xy}$ component to the order parameter of the DDW state breaks the global
TR symmetry, producing a non-zero Kerr signal \cite{Tewari} in conformity
with the experiments \cite{Xia1}. The chiral $d_{xy}+id_{x^{2}-y^{2}}$ ($%
d+id $) density-wave state, as also the regular DDW and the spin density
wave state, has hole and electron pockets as Fermi surfaces in its
excitation spectra. Such reconstructed small Fermi pockets are consistent
with the recently observed quantum oscillation in high magnetic fields in
underdoped YBCO \cite{Doiron,Yelland,Bangura,Jaudet,Chak08}. In this paper,
we discuss an intrinsic anomalous Nernst effect induced by the $d+id$
density-wave state as a direct consequence of the macroscopic TR symmetry
breaking and the presence of the Fermi pockets. Because of the broken TR
symmetry, the ordered state acquires a Berry curvature \cite{Berry} which is
sizable on the Fermi surfaces. It is known that the non-zero Berry curvature
can produce the anomalous Hall \cite{Taguchi,Lee1,Jung,Fang} and Nernst \cite%
{Lee2,Xiao} effects in ferromagnets. We focus here on the anomalous Nernst
effect for the high-$T_{c}$ cuprates, because the corresponding coefficient
has been extensively measured \cite{Xu,Wang1,Wang2}.

Nernst signal for unconventional density waves, such as the DDW state,  was
studied earlier in Ref.~[\onlinecite{Miak}] and references therein. In these
papers, however,  the order parameter of the bare DDW state was used, for
which the Nernst effect was induced by the external magnetic field. On the
other hand, in the present paper we consider the superposition of two
different $d$-wave order parameters (motivated by the PKE measurements \cite%
{Xia1, Tewari}), and the spontaneous breakdown of time reversal symmetry
leads to the Berry curvature, which acts as a magnetic field.  Estimating
the degree of TR symmetry breaking from the PKE measurements of Ref.~\cite%
{Xia1}, 
we calculate the expected anomalous Nernst signal in the underdoped phase of
YBCO near $T^{\ast }$. We stress that even though we model the pseudogap by
a chiral DDW state, the basic conclusions are more robust; the broken TR
symmetry and well-defined Fermi surfaces, both of which have now been
experimentally verified, necessarily imply the anomalous Nernst effect which
should be observable. Note that, recent neutron scattering experiments \cite%
{Mock,Fauque} have appeared to indicate a TR breaking state without
translational symmetry breaking in the pseudogap regime \cite{Varma}. We
expect an anomalous Nernst effect for such a state as well, if it breaks the
TR symmetry globally.

\section{Berry curvature of the chiral DDW state}

The order parameter of the $d_{xy}+id_{x^{2}-y^{2}}$ density wave state \cite%
{Nayak} is a combination of two density waves with different angular
patterns
\begin{equation}
\left\langle c_{\mathbf{k+Q}\alpha }^{\dagger }c_{\mathbf{k}\beta
}\right\rangle =\left( \Delta _{\mathbf{k}}+iW_{\mathbf{k}}\right) \delta
_{\alpha \beta },  \label{Order-Parameter}
\end{equation}%
where $c^{\dagger },c$ are the electron creation and annihilation operators
on the 2D square lattice of Copper atoms, $\mathbf{k}$ is a 2D momentum, $%
\mathbf{Q}$ is the momentum space modulation vector $(\pi ,\pi )$, and $%
\alpha $, $\beta $ are the spin indices. $W_{\mathbf{k}}=\frac{W_{0}}{2}%
(\cos k_{x}-\cos k_{y})$ and $\Delta _{\mathbf{k}}=-\Delta _{0}\sin
k_{x}\sin k_{y}$ are the order parameter amplitudes of the $id_{x^{2}-y^{2}}$
and $d_{xy}$ density wave components, respectively. The imaginary part, $iW_{%
\mathbf{k}}$, of the order parameter breaks the microscopic TR symmetry
giving rise to spontaneous currents along the nearest neighbor bonds of the
square lattice. The spontaneous currents produce a staggered magnetic flux,
which averages to zero on the macroscopic scale. The $d_{xy}$ component of
the density wave, $\Delta _{\mathbf{k}}$, leads to the staggered modulation
of the diagonal electron tunneling between the next-nearest neighbor lattice
sites. Such staggered modulation breaks the symmetry between the plaquettes
with positive and negative circulation and, thus, breaks the macroscopic TR
symmetry. Such macroscopic TR symmetry breaking may account for the nonzero
PKE observed in the recent experiments \cite{Xia1, Tewari}.

The Hartree-Fock Hamiltonian appropriate for the mean-field $d+id$ density
wave is given by
\begin{equation}
H=\sum_{\mathbf{k}\in RBZ}\Psi _{\mathbf{k}}^{+}\left(
\begin{array}{cc}
\varepsilon _{\mathbf{k}}-\mu & D_{\mathbf{k}}\exp \left( i\theta _{\mathbf{%
\ k}}\right) \\
D_{\mathbf{k}}\exp \left( -i\theta _{\mathbf{k}}\right) & \varepsilon _{%
\mathbf{k}+\mathbf{Q}}-\mu%
\end{array}%
\right) \Psi _{\mathbf{k}},  \label{Hamiltonian}
\end{equation}%
where $\Psi _{\mathbf{k}}^{+}=\left(
\begin{array}{cc}
c_{\mathbf{k}}^{\dag } & c_{\mathbf{k}+\mathbf{Q}}^{\dag }%
\end{array}%
\right) $, $\varepsilon _{\mathbf{k}}$ is the free electron band structure, $%
\varepsilon _{\mathbf{k}}=-2t(\cos k_{x}+\cos k_{y})+4t^{\prime }\cos
k_{x}\cos k_{y}\label{free}$, and $\mu $ is the chemical potential. The
order parameter has been rewritten as $D_{\mathbf{k}}\exp \left( i\theta _{%
\mathbf{k}}\right) $ with the amplitude $D_{\mathbf{k}}=\sqrt{W_{\mathbf{k}%
}^{2}+\Delta _{\mathbf{k}}^{2}}$ and the phase $\theta _{\mathbf{k}}=\pi
\Theta \left( -\Delta _{\mathbf{k}}\right) +\arctan \left( W_{\mathbf{k}%
}/\Delta _{\mathbf{k}}\right) $, where $\Theta \left( x\right) $ is the step
function. In writing the Hamiltonian, the first Brillouin zone has been
folded to the magnetic or reduced Brillouin zone (RBZ)
to treat the $\mathbf{Q}=(\pi ,\pi )$ modulation effectively. The energy
spectrum of the Hamiltonian (\ref{Hamiltonian}) contains two bands with
eigenenergies\ $E_{\pm }\left( \mathbf{k}\right) =w_{0}\pm w\left( \mathbf{k}%
\right) $, where $w_{0}\left( \mathbf{k}\right) =-\mu +\left( \varepsilon _{%
\mathbf{k}}+\varepsilon _{\mathbf{k}+\mathbf{Q}}\right) /2$, $w\left(
\mathbf{k}\right) =\sqrt{F_{\mathbf{k}}^{2}+D_{\mathbf{k}}^{2}}$ with $F_{%
\mathbf{k}}=\left( \varepsilon _{\mathbf{k}}-\varepsilon _{\mathbf{k}+%
\mathbf{Q}}\right) /2$.

\begin{figure}[t]
\includegraphics[scale=0.45]{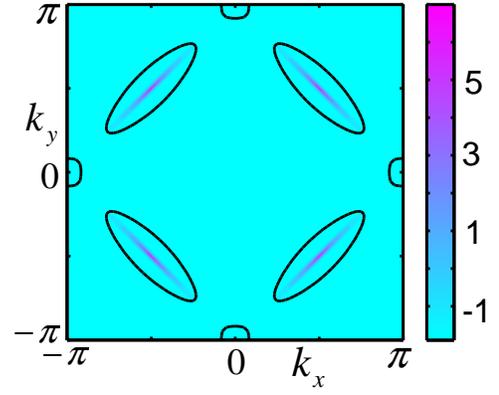}
\caption{(Color online) Logarithm of the Berry curvature $\Omega _{+}\left(
\mathbf{k}\right) $ plotted on the Brillouin zone. The Berry curvature is
sharply peaked at the points ($\pm \frac{\protect\pi }{2},\pm \frac{\protect%
\pi }{2}$). The ellipses and the half-circles are the hole and the electron
pockets of the $d+id$ state, respectively. $t=0.3$ eV, $t^{\prime }=0.09$
eV, $\protect\mu =-0.26$ eV, $W_{0}=0.08$ eV, $\Delta _{0}=0.004$ eV. }
\label{bc}
\end{figure}
Berry phase is a geometric phase acquired by the wavefunction when the
Hamiltonian of a physical system undergoes transformation along a closed
contour in the parameter space \cite{Berry}. For the $d+id$ Hamiltonian (\ref%
{Hamiltonian}), the relevant parameter space is the space of the crystal
momentum $\mathbf{k}$. The eigenfunctions of the Hamiltonian are therefore $%
\mathbf{k}$-dependent, and the overlap of two wavefunctions infinitesimally
separated in the $\mathbf{k}$-space defines the Berry phase connection $%
\mathbf{A}_{\mathbf{k}}=\left\langle \Phi _{n}^{\dagger }\left( \mathbf{k}%
\right) \right\vert i\mathbf{\nabla }_{\mathbf{k}}\left\vert \Phi _{n}\left(
\mathbf{k}\right) \right\rangle $, where $\Phi _{n}\left( \mathbf{k}\right) $
is the periodic amplitude of the Block wave function, and $n$ is the band
index. The Berry phase connection corresponds to an effective vector
potential in the momentum space, and its line integration around a close
path gives the Berry phase. The Berry curvature, the Berry phase per unit
area in the $\mathbf{k}$ space, is given by
\begin{equation}
\mathbf{\Omega }_{n}\left( \mathbf{k}\right) =\mathbf{\nabla }_{\mathbf{k}%
}\times \mathbf{A}_{\mathbf{k}},\quad \mathbf{A}_{\mathbf{k}}=\left\langle
\Phi _{n}^{\dagger }\left( \mathbf{k}\right) \right\vert i\mathbf{\nabla }_{%
\mathbf{k}}\left\vert \Phi _{n}\left( \mathbf{k}\right) \right\rangle .
\label{Berry}
\end{equation}%
The Berry curvature $\mathbf{\Omega }_{n}\left( \mathbf{k}\right) $, thus,
acts as an effective magnetic field in the momentum space and enters in the
equations of motion of the wavepacket. For a system invariant under both
time reversal and spatial inversion symmetries, the Berry curvature $\mathbf{%
\Omega }_{n}\left( \mathbf{k}\right) =0$ for every $\mathbf{k}$. However,
the $d+id$ density-wave state breaks the macroscopic TR symmetry, therefore
the Berry curvature can acquire non-zero values.

To calculate the non-zero Berry curvature, we find the eigenstates of the
Hamiltonian (\ref{Hamiltonian}), which are given by $\Phi _{\pm }\left(
\mathbf{k}\right) =\left( u_{\pm }\left( \mathbf{k}\right) e^{i\theta _{%
\mathbf{k}}/2},v_{\pm }\left( \mathbf{k}\right) e^{-i\theta _{\mathbf{k}%
}/2}\right) $, where $+$ and $-$ correspond to the upper and lower bands
with the energy dispersion $E_{+}\left( \mathbf{k}\right) $ and $E_{-}\left(
\mathbf{k}\right) $, respectively. The coefficients $u_{\pm }\left( \mathbf{k%
}\right) $ and $v_{\pm }\left( \mathbf{k}\right) $ in the eigenstates $\Phi
_{\pm }\left( \mathbf{k}\right) $ are straightforwardly obtained from the
matrix (\ref{Hamiltonian}). Substituting the eigenstates $\Phi _{\pm }\left(
\mathbf{k}\right) $ into Eq. (\ref{Berry}), we find $\mathbf{\Omega }_{\pm
}\left( \mathbf{k}\right) =-\frac{1}{2}\mathbf{\nabla }_{\mathbf{k}}\left(
u_{\pm }^{2}\left( \mathbf{k}\right) -v_{\pm }^{2}\left( \mathbf{k}\right)
\right) \times \mathbf{\nabla }_{\mathbf{k}}\theta _{\mathbf{k}}$. In the
pure DDW state, $\theta _{\mathbf{k}}=\pi /2$ is a constant, therefore $%
\mathbf{\Omega }_{\pm }\left( \mathbf{k}\right) =0$ and there are no Berry
phase effects. However, in the $d+id$ density-wave state, the phase $\theta
_{\mathbf{k}}=\pi \Theta \left( -\Delta _{\mathbf{k}}\right) +\arctan \left(
W_{\mathbf{k}}/\Delta _{\mathbf{k}}\right) $ depends on the values of order
parameters $W_{k}$ and $\Delta _{k}$, and can vary in the $\mathbf{k}$
space, therefore $\mathbf{\Omega }_{\pm }\left( \mathbf{k}\right) $ can
acquire nonzero values. Because the momentum $\mathbf{k}$ is restricted to
the $xy$ plane, only the $z$ component of $\mathbf{\Omega }_{\pm }\left(
\mathbf{k}\right) $ can be nonzero, which we will denote as $\Omega _{\pm
}\left( \mathbf{k}\right) $. After some straightforward algebra, we find
\begin{eqnarray}
\Omega _{\pm }\left( \mathbf{k}\right)  &=&\mp \frac{1}{2w^{3}\left( \mathbf{%
k}\right) }\mathbf{w}_{k}\cdot \left[ \frac{\partial \mathbf{w}_{k}}{%
\partial k_{x}}\times \frac{\partial \mathbf{w}_{k}}{\partial k_{y}}\right] ,
\label{BC3} \\
&=&\pm \frac{t\Delta _{0}W_{0}}{w^{3}\left( \mathbf{k}\right) }\left( \sin
^{2}k_{y}+\cos ^{2}k_{y}\sin ^{2}k_{x}\right)   \label{BC4}
\end{eqnarray}%
where $\mathbf{w}_{\mathbf{k}}$ is a three-component vector, $\mathbf{w}_{%
\mathbf{k}}=\left( -\Delta _{\mathbf{k}},-W_{\mathbf{k}},F_{\mathbf{k}%
}\right) $, and it enters into the Hamiltonian density in (\ref{Hamiltonian}%
) as $\hat{H}=w_{0}\hat{I}+\mathbf{w}_{\mathbf{k}}\cdot \mathbf{\hat{\tau}%
_{i}}$. Here $\mathbf{\hat{\tau}_{i}}$ $(i=1,2,3)$ are the Pauli matrices
and $\hat{I}$ is the $2\times 2$ unit matrix operating on the spinors $\Psi
_{\mathbf{k}}^{+}$, $\Psi _{\mathbf{k}}$. We see from Eq. (\ref{BC4}) that
the Berry curvature is nonzero only when the amplitudes $\Delta _{0}$ and $%
W_{0}$ of the $d_{xy}$ and $id_{x^{2}-y^{2}}$ order parameters are both
nonzero. The Berry curvatures have opposite signs in the upper and the lower
bands: $\Omega _{+}(\mathbf{k})=-\Omega _{-}(\mathbf{k})$. In Fig. \ref{bc},
we plot the Berry curvature $\Omega _{+}$ with respect to the momentum $%
\mathbf{k}$ for a set of parameters in the $d+id$ state. We see that $\Omega
_{+}$ peaks at $\left( \pm \frac{\pi }{2},\pm \frac{\pi }{2}\right) $, where
$w\left( \mathbf{k}\right) $ reaches the minimum and the corresponding
points in the $\mathbf{k}$ space are the points of near degeneracy between
the two bands. The value of $\Omega _{+}$ decreases dramatically along slim
ellipses whose long axes lay on the RBZ boundary lines $k_{y}\pm k_{x}=\pm
\pi $ where $w\left( \mathbf{k}\right) $ and the band splitting are the
smallest. The peaks of the Berry curvature $\Omega _{\pm }$ correspond to
magnetic monopoles in the momentum space \cite{Fang}.

\section{Anomalous Nernst effect in the chiral DDW state}

In the experiments to observe Nernst effect \cite{Lee2,Wang1,Wang2}, a
temperature gradient $-\mathbf{\nabla }T$, applied along, say, the $\hat{x}$
direction produces a measurable transverse electric field. The charge
current along $\hat{x}$ driven by $-\mathbf{\nabla }T$ is balanced by a
backflow current produced by an electric field $\mathbf{E}$. The total
charge current in the presence of $\mathbf{E}$ and $-\mathbf{\nabla }T\,$ is
thus given by, $J_{i}=\sigma _{ij}E_{j}+\alpha _{ij}\left( -\partial
_{j}T\right) $, where $\sigma _{ij}$\ and $\alpha _{ij}$ are the electric
and the thermoelectric conductivity tensors, respectively. In the
experiments, $\mathbf{J}$ is set to zero and the Nernst signal, defined as
\begin{equation}
e_{N}\equiv E_{y}/\left\vert \mathbf{\nabla }T\right\vert =\rho \alpha
_{xy}-S\tan \theta _{H},  \label{Nsignal}
\end{equation}%
is measured, where $\alpha _{xy}$ is Nernst conductivity defined via the
relation $J_{x}=\alpha _{xy}\left( -\partial _{y}T\right) $ in the absence
of the electric field, $\rho =1/\sigma _{xx}$ is the longitudinal
resistance, $S=E_{x}/\left\vert \mathbf{\nabla }T\right\vert =\rho \alpha
_{xx}$ is the thermopower, and $\tan \theta _{H}=\sigma _{xy}/\sigma _{xx}$
is the Hall angle. For a relatively modest hole concentration away from the
severely underdoped regime in the cuprates, the second term in Eq.~(\ref%
{Nsignal}) is experimentally observed to be small \cite{Wang1}. As long as
the second term is small, $\rho \alpha _{xy}$ completely defines the Nernst
signal, but in the most general case one should extract $\alpha _{xy}$ from
the experimental data, as it was done in Refs.~\cite{Lee2, Wang1}, to
compare with our theory.

The Berry-phase effects have found much success in explaining the anomalous
Hall and Nernst effects in ferromagnets \cite{Taguchi,Fang,Lee2,Xiao}. In
the presence of an external electric field $\vec{E}$ along the $\hat{x}$
direction, the anomalous Hall current is along the transverse $\hat{y}$
direction. The anomalous DC Hall conductivity is found to be
\begin{equation}
\sigma _{xy}=-\frac{e^{2}}{\hbar }\int_{\text{RBZ}}\frac{dk_{x}dk_{y}}{%
\left( 2\pi \right) ^{2}}\Omega _{-}\left[ f\left( E_{-}\left( k\right)
\right) -f\left( E_{+}\left( k\right) \right) \right] ,  \label{DCHall}
\end{equation}%
where $f\left( E_{n}\right) =1/\left( 1+\exp \left( \beta E_{n}\right)
\right) $ is the Fermi distribution function at a temperature $T$, $\beta
=1/k_{B}T$ and we have used $\Omega _{+}=-\Omega _{-}$. Eq. (\ref{DCHall})
agrees with the DC Hall conductivity of the $d+id$ density wave obtained
earlier using a different approach \cite{Tewari,Yakovenko,Littlewood}. For
half filling ($t^{\prime }=0$, $\mu =0$), when the system is a band
insulator, its value is quantized, $e^{2}/2\pi \hbar $ \cite%
{Yakovenko,Littlewood} per spin component. It changes continuously as the
system deviates from half filling and the Fermi pockets appear \cite{Tewari}%
. We will see below that the anomalous Nernst effect is zero in the case of
half filling and becomes non-zero only when there are hole and electron
pockets in the spectrum.

\begin{figure}[t]
\includegraphics[scale=0.45]{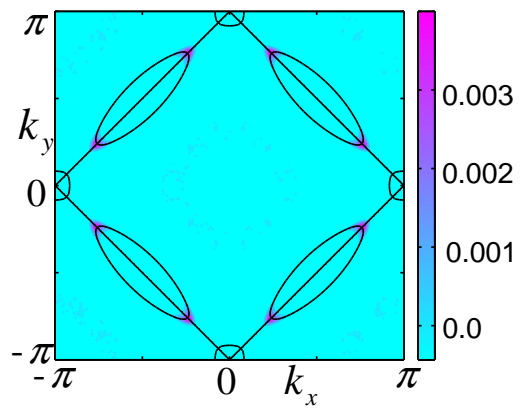}
\caption{(Color online) The contour plot of the integrand of Eq. (\protect
\ref{Nernst}) for the anomalous Nernst conductivity. The plotting parameters
are the same as those in Fig. 1 except $\Delta _{0}=0.0008$ eV. $T=130$ K.
The main contribution to the anomalous Nernst signal comes from the lower
band (hole pockets). The upper band (electron pockets) gives a negligible
contribution.}
\label{integrand}
\end{figure}

In order to obtain the coefficient $\alpha _{xy}$, it is more convenient to
calculate the coeffecient $\bar{\alpha}_{xy}$, which determines the
transverse heat current $\mathbf{J}^{h}$ in response to the electric field $%
\mathbf{E}$: $J_{x}^{h}=\bar{\alpha}_{xy}E_{y}$. It is related to $\alpha
_{xy}$ by the Onsager relation $\bar{\alpha}_{xy}=T\alpha _{xy}$ \cite%
{Xiao,Cooper}. In the presence of the Berry curvature and the electric
field, the electron velocity acquires the additional anomalous term $\hbar
\mathbf{v_{\mathbf{k}}}=e\mathbf{E}\times \mathbf{\Omega }(\mathbf{k})$ \cite%
{Lee2, Xiao}. Multiplying this velocity by the entropy density of the
electron gas, we obtain the coefficient for the transverse heat current:
\begin{equation}
\bar{\alpha}_{xy}=T\alpha _{xy}=\frac{e}{\beta \hbar }\sum_{n=\pm }\int_{%
\text{RBZ}}\frac{dk_{x}dk_{y}}{\left( 2\pi \right) ^{2}}\,\Omega _{n}(%
\mathbf{k})\,s_{n}(\mathbf{k}).  \label{entropy}
\end{equation}%
Here $s(\mathbf{k})=-f_{\mathbf{k}}\ln f_{\mathbf{k}}-(1-f_{\mathbf{k}})\ln
(1-f_{\mathbf{k}})$ is the entropy density of the electron gas, $f_{\mathbf{k%
}}=f[E_{n}(\mathbf{k})]$ is the Fermi distribution function, and the sum is
taken over both bands. Using the explicit expression for the Fermi
distribution function, Eq.~(\ref{entropy}) can be transformed to the
following form
\begin{eqnarray}
&&\alpha _{xy}=\frac{e}{\hbar }\frac{1}{T}\sum_{n=\pm }\int_{\text{RBZ}}%
\frac{dk_{x}dk_{y}}{(2\pi )^{2}}\Omega _{n}\times  \label{Nernst} \\
&&\left\{ E_{n}\left( \mathbf{k}\right) f\left( E_{n}\left( \mathbf{k}%
\right) \right) -k_{B}T\log \left[ 1-f\left( E_{n}\left( \mathbf{k}\right)
\right) \right] \right\} .  \nonumber
\end{eqnarray}%
Eq.~(\ref{Nernst}) coincides with the corresponding expression derived in
Ref.~\cite{Xiao} using the semiclassical wavepacket methods and taking into
accound the orbital magnetization of the carriers \cite{Shi}. Relation of
the transverse heat current to the entropy flow was also discussed in Refs.~%
\cite{Cooper} and \cite{Lee2}. At $T=0$, the carrier entropy is zero, $s_{n}(%
\mathbf{k})=0$, so there is no heat current and $\alpha _{xy}=0$. At $T\neq
0 $, we first consider the simple case with $t^{\prime }=0$ and $\mu =0$,
that is, the lower and the upper bands are symmetric with $E_{+}\left(
\mathbf{k}\right) =-E_{-}\left( \mathbf{k}\right) $. It is easy to check
that in this case $\alpha _{xy}=0$ because $\Omega _{+}=-\Omega _{-}$.

In the general case, the entropy $s_{n}\left( \mathbf{k}\right) $ has peaks
on the Fermi surface and decreases dramatically away from the Fermi surface.
Because the Berry curvature $\Omega _{\pm }$ peaks along the RBZ boundary $%
k_{x}\pm k_{y}=\pm \pi $, the integrand in Eq. (\ref{Nernst}) peaks at the
intersections of the Fermi surface and the RBZ boundary, the so called
\textquotedblleft hot spots\textquotedblright\ \cite{Rice,Stojkovic}, which
were shown earlier to be important in the calculations of the Hall
coefficient in the DDW state \cite{Sumanta}. These peaks are clearly seen in
Fig. \ref{integrand}. 

From Eqs. (\ref{DCHall}), (\ref{entropy}), and (\ref{Nernst}), we can show
that, at low temperatures, the Nernst conductivity $\alpha _{xy}$ is related
to the zero temperature Hall conductivity $\sigma _{xy}$ through the Mott
relation \cite{Marder}, which yields
\begin{equation}
\alpha _{xy}=\frac{\pi ^{2}k_{B}^{2}}{3e}\frac{d\sigma _{xy}}{d\mu }T.
\label{Signal}
\end{equation}%
Here the derivative of $\sigma _{xy}$ leads to $\frac{d\sigma _{xy}}{d\mu }=-%
\frac{e^{2}}{\hbar }\int_{\text{RBZ}}\frac{dk_{x}dk_{y}}{\left( 2\pi \right)
^{2}}\Omega _{-}\left[ \delta \left( E_{-}\right) -\delta \left(
E_{+}\right) \right] $. Here $\delta \left( E_{\pm }\right) $ are the delta
functions. Therefore the integrand is nonzero only at the boundary lines of
the hole and electron pockets. In the case of a band insulator ($t^{\prime
}=0$, $\mu =0$) that does not contain the Fermi pockets, $\sigma
_{xy}^{\prime }\left( \mu \right) =0$ and the anomalous Nernst conductivity $%
\alpha _{xy}=0$, even though the DC Hall conductivity, Eq.~(\ref{DCHall}),
is non-zero and, in fact, is quantized \cite{Tewari}. This is because the
quantum Hall current carries no entropy.

\begin{figure}[t]
\includegraphics[width=0.65\linewidth]{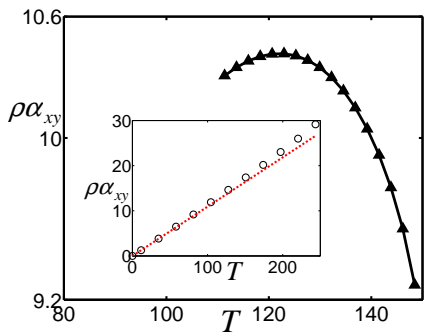}
\caption{(Color online) Plot of the Nernst signal, see Eq.~(\protect\ref%
{Nsignal}), versus temperature $T$. The unit on the y-axis is nV/K. The
inset, with temperature independent order parameters $W_{0}=0.08$ eV, $%
\Delta _{0}=0.1\%W_{0}$ eV, shows that the Nernst signal satisfies the Mott
relation (\protect\ref{Signal}). In the inset, dotted line corresponds to
the Mott relation (\protect\ref{Signal}) and the open circles are obtained
through numerical integration of Eq. (\protect\ref{Nernst}). }
\label{Sig}
\end{figure}

For crude estimate of the Nernst signal, we choose a set of parameters
appropriate for the underdoped YBCO \cite{Anderson}, $t=0.3$ eV, $t^{\prime
}=0.09$ eV, $\mu =-0.26$ eV (corresponding to the hole doping of about
10\%), $d=1.17$ nm (the distance between consecutive 2D layers), $\rho =3$ m$%
\Omega $cm, $W_{0}\left( T\right) =0.1\left( 1-T/T_{W}^{\ast }\right) ^{1/2}$
eV, $\Delta _{0}\left( T\right) =0.0001\left( 1-T/T_{\Delta }^{\ast }\right)
^{1/2}$ eV, and numerically integrate Eq. (\ref{Nernst}), where we made
reasonable assumptions about the transition temperatures, $T_{W}^{\ast
}\approx 150$ K \cite{Xia1} and $T_{\Delta }^{\ast }\approx 250$ K. In Fig.~%
\ref{Sig}, we plot $\rho \alpha _{xy}$ in a temperature regime that is below
$T_{W}^{\ast }$ but much higher than the superconducting transition
temperature $T_{c}\approx 80$ K \cite{Xia1}. Here we have multiplied the
results by 2 to account for the contributions from two spin components. As
temperature drops from $T_{W}^{\ast }$, the order parameter $W_{0}\left(
T\right) $ grows, leading to the increase of the Nernst signal. Close to $%
T_{c}$, the Nernst effect would be dominated by the mobile vortices and our
calculations do not apply there. The estimated value of $\rho \alpha _{xy}$
at $T\sim 130$ K is about $10$ nV/K. This value is about 10\% of the
experimentally observed Nernst signals in underdoped LSCO and BSCCO \cite%
{Wang2} at temperatures much higher than the superconducting $T_{c}$. Note
that the spontaneous Nernst signal discussed above may not be observable
through the DC current measurements \cite{Wang2} without a non-zero magnetic
field because of the macroscopic domains with opposite chiralities present
in a sample at the zero magnetic field.

\section{Conclusion}

In summary, we discuss the non-zero Berry curvature in the $d+id$
density-wave state, which was proposed earlier \cite{Tewari} to explain the
time-reversal symmetry breaking \cite{Xia1} in the pseudogap phase of the
high $T_{c}$ superconductor YBCO. We show that the nonzero Berry curvature,
arising out of the broken time-reversal invariance, and the existence of
Fermi pockets in the cuprates directly imply an anomalous Nernst effect
which should be measurable. We note that measurable Nernst signals have been
found in underdoped LSCO and BSCCO \cite{Wang2} even at temperatures much
higher than $T_{c}$, and we propose that a TRS breaking state, such as the
chiral DDW state, may be the origin of these signals. The anomalous Nernst
effect at the pseudogap temperatures will constitute a further proof of an
ordered state, with broken time-reversal invariance, to be responsible for
the pseudogap phenomena in the cuprates.

This work is supported by ARO-DARPA and LPS-CMTC.

\end{document}